\documentclass[aps,prl,reprint,showpacs,showkeys,preprintnumbers,superscriptaddress,nofootinbib]{revtex4-2}

\usepackage{graphicx,epsfig,color,xcolor}

\usepackage[english]{babel}
\usepackage{amssymb,amsfonts,amsmath,physics}
\usepackage{mathtools}
\usepackage{siunitx}
\usepackage{verbatim}
\usepackage{ulem}
\usepackage{soul}
\usepackage{lipsum, babel}
\usepackage{acronym}
\usepackage{aas_macros}

\usepackage[colorlinks=true
,urlcolor=DARKBLUE
,anchorcolor=DARKBLUE
,citecolor=DARKBLUE
,filecolor=DARKBLUE
,linkcolor=DARKBLUE
,menucolor=DARKBLUE
,pagecolor=DARKBLUE
,linktocpage=true
,pdfproducer=medialab
,pdfa=true
,bookmarks=true
]{hyperref}

\DeclareMathOperator{\sinc}{sinc}

\newcommand{\be}{\begin{equation}} \newcommand{\ee}{\end{equation}}
\newcommand{\bea}{\begin{eqnarray}} \newcommand{\eea}{\end{eqnarray}}

\newcommand{\newsec}[1]{\textbf{#1.}}

\definecolor{MONZA}{HTML}{CF000F}
\definecolor{DARKBLUE}{HTML}{00008b}
\definecolor{DARKMAGENTA}{HTML}{8b008b}
\definecolor{DARKCYAN}{HTML}{008B8B}
\definecolor{DARKORANGE}{HTML}{FF8C00}
\definecolor{OBSERVATORY}{HTML}{049372}
\definecolor{GREENBAMBOO}{HTML}{006442}
\definecolor{TURQUOISE}{HTML}{36D7B7}
\definecolor{JUNGLEGREEN}{HTML}{26C281}

\begin{document}

\title{Non-spherical effects on the mass function of Primordial Black Holes}

\author{Albert Escrivà}
\email{escriva.manas.alberto.k0@f.mail.nagoya-u.ac.jp}
\affiliation{\mbox{Division of Particle and Astrophysical Science, Graduate School of Science,} \\ Nagoya University, Nagoya 464-8602, Japan}
\affiliation{\mbox{Division of Science, National Astronomical Observatory of Japan,} \\Mitaka, Tokyo 181-8588, Japan}
\affiliation{\mbox{Institute for Advanced Research, Nagoya University}, \\
Furo-cho Chikusa-ku, Nagoya 464-8601, Japan} 

\author{Chul-Moon Yoo}
\email{yoo.chulmoon.k6@f.mail.nagoya-u.ac.jp}
\affiliation{\mbox{Division of Particle and Astrophysical Science, Graduate School of Science,} \\ Nagoya University, Nagoya 464-8602, Japan}
\affiliation{\mbox{Kobayashi Maskawa Institute,} 
Nagoya University, Nagoya 464-8602, Japan}
\begin{abstract}
In this letter, we investigate the impact of non-spherical effects on the Primordial Black Hole mass function, based on the ellipticity-dependent threshold calculated by performing $3+1$ relativistic numerical simulations. We consider an equation of state of radiation $w:=P/\rho=1/3$ and a softer one  $w=1/10$ with $P$ and $\rho$ being the pressure and energy density, respectively. We suppose that the curvature perturbations obey Gaussian statistics with a monochromatic power spectrum and examine the most probable ellipsoidal configurations utilizing peak theory. We also suppose the critical scaling law of the PBH mass near the threshold following the known results. The simulations \cite{companion} show that the non-sphericity can easily prevent the system from black hole formation when the initial fluctuation amplitude is near the threshold (critical scaling regime). Nevertheless, 
we show that the non-spherical effects make the mass function just a few times smaller and are insignificant on the mass function distribution, including the power-law scaling in the small mass region.
\end{abstract}
\pacs{
}

\maketitle
\acresetall

\acrodef{GW}{gravitational wave}
\acrodef{PT}{phase transition}
\acrodef{SC}{smooth crossover}
\acrodef{SM}{Standard Model}
\acrodef{QCD}{Quantum Chromodynamics}
\acrodef{EW}{electroweak}
\acrodef{CMB}{cosmic microwave background}
\acrodef{PBH}{primordial black hole}
\acrodef{DM}{Dark Matter}
\acrodef{FLRW}{Friedmann‐-Lema\^itre--Robertson--Walker}
\acrodef{MS}{Misner--Sharp}

\newsec{Introduction}
Primordial Black Holes (PBHs) \cite{Zeldovich:1967lct,10.1093/mnras/152.1.75,Carr:1974nx} are black holes that could have been formed in the early Universe without a stellar origin through various mechanisms (see \cite{Escriva:2022duf} for a detailed list of mechanisms together with reviews on broad perspectives on PBHs \cite{Khlopov:2008qy,Sasaki:2018dmp,Carr:2020gox,Green:2020jor,Carr:2020xqk,Escriva:2022duf}). 
%
The most extensively considered scenario is the collapse of super-horizon curvature fluctuations, particularly during a radiation-dominated era \cite{Carr:1975qj}. PBHs remain a remarkable candidate to explain the dark matter of a significant fraction of it \cite{Chapline:1975ojl}, in particular in the asteroid mass range with $M_{\rm PBH} \in [10^{-15}, 10^{-10}] M_{\odot}$.

A common approach to studying the PBH formation process is the assumption of spherical symmetry (see \cite{Escriva:2021aeh} for a review focusing on numerical results), which means that we assume that the initial super-horizon curvature profiles (generated during inflation) are spherically symmetric and, therefore, the gravitational collapse is spherical when these fluctuations reenter the cosmological horizon. 
This assumption is based on the fact that according to peak theory \cite{1986ApJ...304...15B}, large peaks (those responsible for producing a sizable number of PBHs) tend to be almost spherically symmetric, and the possible deviation from sphericity is small according to the probability distribution of the ellipsoidal shape of the curvature fluctuations. Nevertheless, the validity of this assumption has not been tested, and indeed the typical curvature fluctuation shape is ellipsoidal. In specific inflationary models, deviations from sphericity can be observed, as shown in \cite{Mizuguchi:2024kbl} using stochastic lattice simulations. On the other hand, non-spherical effects are known to have a significant impact during a dust-dominated epoch \cite{1965ApJ...142.1431L,KHLOPOV1980383}.

In estimating the abundance of PBHs during the radiation-dominated epoch of our Universe, 
it has been essential to develop precise statistical methods to account for PBH production. 
Peak theory \cite{1986ApJ...304...15B} 
provides
the current state-of-the-art method, where different procedures have been proposed relying on the assumption of spherical symmetry (see \cite{Yoo:2022mzl,Germani:2023ojx,Young:2024jsu} for reviews). 
The estimation of PBH abundance is highly sensitive 
to the threshold for black hole formation. 
The threshold value 
has been extensively studied through numerical simulations assuming spherical symmetry \cite{Shibata:1999zs,Niemeyer:1999ak,Musco:2004ak,Musco:2008hv,Nakama:2013ica,Nakama:2014fra,Harada:2015yda,Escriva:2019nsa,Musco:2018rwt,Escriva:2019phb,Escriva:2020tak,Yoo:2021fxs,Escriva:2022yaf,Escriva:2023qnq,Uehara:2024yyp} but there are only a few works on non-spherical cases~\cite{Yoo:2020lmg,companion}.
To estimate the PBH mass function, we need to estimate the resultant black hole mass 
through the gravitational collapse. 
In spherically symmetric cases, it is known that a critical scaling law is universally realized 
for a one-parameter family in the phase space~\cite{Choptuik:1992jv,Koike:1995jm,Maison:1995cc}. 
The formation of primordial black holes is not an exception~\cite{Niemeyer:1999ak,IHawke_2002,Musco:2008hv}. 
More specifically, in the scaling regime, we find 
\begin{equation}
M \propto (\mu-\mu_{\rm c,sp})^{\gamma}, 
\label{eq:scalingM}
\end{equation}
where $\mu$ is the initial amplitude, and $\mu_{\rm c,sp}$ and $\gamma$ 
are the threshold value of $\mu$ and the critical exponent, respectively.
While the threshold value $\mu_{\rm c,sp}$ is profile dependent, 
the value of the critical exponent 
is known to be universal for any one-parameter sequence of initial conditions
under the spherical symmetry with a perfect fluid being the dominant component. 
A characteristic scaling law for the mass function has been predicted based 
on this scaling behavior~\cite{Niemeyer:1997mt,Yokoyama:1998xd,Kitajima:2021fpq}.

In the context of non-spherical collapse, the relevance of the scaling behaviour 
has not been 
completely
addressed 
yet, but previous studies \cite{Gundlach:1999cw,Baumgarte:2015aza,Celestino:2018ptx} for the collapse of perfect fluid on asymptotically flat spacetime suggest that the power-law relation may be preserved with non-sphericities 
depending on 
the equation of state of the perfect fluid (see Ref.~\cite{Chiba:2017rvs} for one of its application). 

The impact of non-sphericities on PBH abundance and mass function was explored in \cite{Kuhnel:2016exn} 
(see also \cite{Harada:2016mhb}) 
based on the Carr's formula~\cite{1974ApJ...187..425P,Carr:1975qj} for the abundance. 
This study incorporated assumptions about how the collapse threshold varies with aspherical configurations based on halo collapse models \cite{Sheth:1999su,Angrick:2010qg}. The results indicated that the PBH mass function could be significantly suppressed when non-sphericities are considered, depending on the sensitivity of the threshold to these effects. However, 
it is clear that non-spherical simulations are necessary for more reliable predictions. 

In this work, we aim to clarify
whether non-spherical effects can significantly alter the mass function of PBHs. 
To achieve this, we combine results from $3+1$ PBH formation simulations \cite{companion} with a precise determination of PBH abundance using peak theory \cite{1986ApJ...304...15B}, 
taking the likelihood of non-sphericity parameters into account. 
Our study focuses on a monochromatic power spectrum 
assuming the random Gaussian statistics of the curvature fluctuations, while our results may also directly apply to cases with a sharply-peaked spectrum. 

\newsec{Non-spherical peaks and the threshold}



We assume that the Universe is filled by a perfect fluid with a linear equation of state $P= w \rho$, 
being $P$ the pressure and $\rho$ the energy density, considering the cases $w=1/10$ and $w=1/3$. 
Then, we consider super-horizon curvature fluctuations that are initially frozen at super-horizon scales, characterized by a monochromatic power spectrum $\mathcal{P}_{\zeta}(k) = \mathcal{A}_{\zeta} \, \delta(\ln(k/k_p))$ peaked at a wave mode $k_p$. We fix the initial condition ensuring that $k_p/(a H_b) \ll 1$ to fulfill the super-horizon condition, 
where $a$ and $H_b$ are the scale factor and Hubble expansion rate 
for the Friedmann-Lema\^itre-Robertson-Walker background. 
According to peak theory \cite{1986ApJ...304...15B} with a monochromatic spectrum, the typical profile (see (7.8) in Ref.~\cite{1986ApJ...304...15B}) 
characterized by the wave mode $k_p$, peak amplitude $\eval{\bar{\zeta}}_{r=0}=\mu$, ellipticity $e$, and prolateness $p$ 
is given by 
 \begin{align}
 \label{eq:non_spherical_zeta}
     \bar{\zeta}(x,y,z;\mu) &=  \bar{\zeta}_{\rm sp}(r;\mu)  \nonumber\\
     &+ \frac{5\mu}{2k_p^3r^3}\left( 3 k_p r \cos(k_p r)+(r^3 k^3_p -3) \sin(k_p r) \right) \nonumber\\
     &\times \left[\frac{3 e}{r^2} (z^2-y^2)+p\left( 1-3\left(\frac{x}{r}\right)^2  \right) \right],  
 \end{align}
 where $r=\sqrt{x^2+y^2+z^2}$ with $x$, $y$, and $z$ being Cartesian coordinates, and 
 $\bar{\zeta}_{\rm sp}(r;\mu)=\mu \sinc(k_p \, r)$ is the typical profile for the spherical case. 

\begin{figure} 
    \centering
    \includegraphics[width=0.45\textwidth]{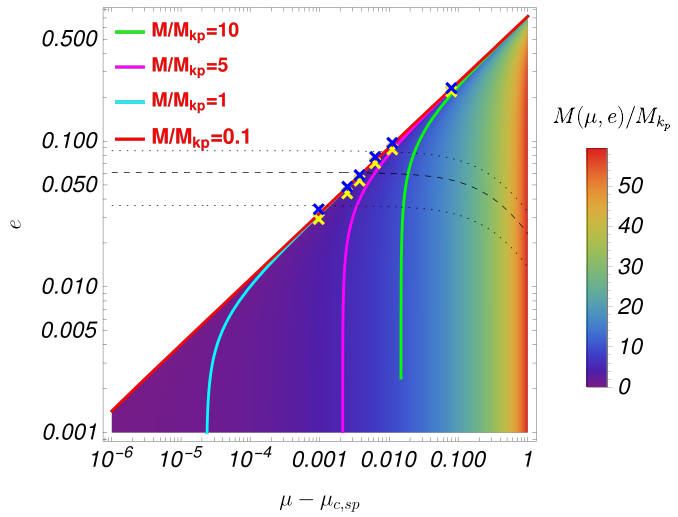}
    \caption{Density plot of the PBH mass 
    in the parameter space of 
 $e$ and $\mu-\mu_{\rm c,sp}$. 
 The dashed line indicates the mean value of $e$ with one sigma deviation specified by the dotted lines (see \cite{companion} for the analytical equation). 
 The solid colored lines indicate the lines of constant mass. 
 The cross symbols indicate the cases with PBH formation (yellow) and non-formation (blue) from the numerical results obtained in \cite{companion}. 
 The case shown corresponds to radiation fluid $w=1/3$.} 
    \label{fig:non_spherical_mass}
\end{figure}

In the spherically symmetric case, we find the threshold $\mu_{\rm c,sp}$ of $\mu$ for PBH formation. 
That is, PBH forms for $\mu$ greater than $\mu_{\rm c,sp}$ with $e=p=0$. 
In the companion paper \cite{companion}, we found that, for the value of $\mu$ slightly above the threshold $\mu_{\rm c,sp}$, we find the threshold $e_{\rm c}$ dividing the parameter region of $e$ into domains of PBH formation and non-formation 
with $p=0$. 
More specifically, we found that, for the value of $e$ greater than $e_{\rm c}$, no PBH formation is found, where $e_{\rm c}$ is approximately given as a function of $\mu$ as 
\begin{equation}
    e_{\rm c}(\mu) = \mathcal{K}_e \left(\frac{\mu - \mu_{\rm c,sp}}{\mu_{\rm c,sp}}\right)^{\gamma_e}, 
    \label{eq:ec}
\end{equation}
with 
$\mathcal{K}_e \approx 0.532$ and $\gamma_e \approx 0.413$ for $w=1/10$, and $\mathcal{K}_e \approx 0.563$ and $\gamma_e \approx 0.448$ for $w=1/3$ (see Fig.~\ref{fig:non_spherical_mass}).

Here we set $p=0$ because the probability distribution of $p$ is maximized at $p \approx 0$, 
and the dominant contribution comes from the cases with $p\approx0$ as will be explicitly shown later. Further details on the effect of the non-vanishing value of $p$ are 
given in the companion paper~\cite{companion}.

\begin{figure} 
    \centering 
    \includegraphics[width=0.45\textwidth]{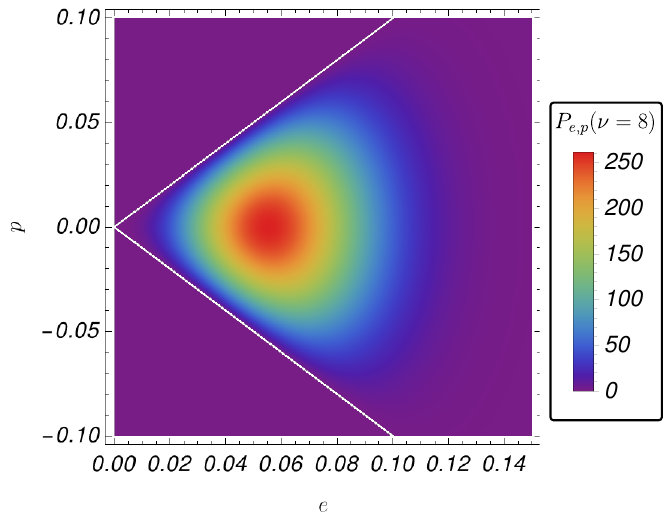}
    \caption{Probability distribution of the prolateness $p$ and ellipticity $e$ of the shape fixing $\nu=8$ given by Eq.~\eqref{eq:prob_e_p}. For large peaks, only a small region in the parameter space $(e,p)$ has a non-negligible probability with $\langle e \rangle \approx  6 \cdot 10^{-2}, \langle p \rangle \approx 10^{-3}$.}
    \label{fig:P_ep}
\end{figure}

Equation \eqref{eq:ec} indicates that, for the value of $\mu$ very close to $\mu_{\rm c,sp}$, 
the possible parameter region of $e$ for PBH formation is limited to the tiny region (see Fig.\ref{fig:P_ep}). 
Naively combining this result with the expression \eqref{eq:scalingM}, 
one may conclude that the mass function is significantly suppressed 
in the small mass scaling region
by 
the effect of ellipticity (see Fig.14 in \cite{companion}). 
However, this is not necessarily the case because the relation between the mass and 
$\mu$ should be dependent on the ellipticity $e$.

Some numerical studies have examined the critical collapse of a perfect fluid in asymptotically flat spacetime with aspherical initial data \cite{Baumgarte:2015aza,Celestino:2018ptx} (see also \cite{Baumgarte:2016xjw,Gundlach:2016jzm,Gundlach:2017tqq} for the case of rotating fluids). 
Those works suggested that the characteristic power-law scaling for $M$ remains valid. 
At least in their settings, the deviations from sphericity in the initial data 
damps down in the time evolution\footnote{The damping of the nonsphericity can also be observed in our numerical simulations shown in 
\cite{companion}, where 
oscillation of the ellipsoidal shape during the collapse
leads to a final state with a spherical shape. }, and do not break the critical scaling law\footnote{These findings are consistent with the pioneering perturbative analysis in \cite{Gundlach:1999cw}, which showed that for $0.11 \lesssim w \lesssim 0.49$, non-spherical modes in the initial conditions decay 
recovering the critical solution in 
spherical symmetry. 
For stiffer or softer values of $w$ outside this range, 
the non-spherical modes may not decay
and break
the scaling law behaviour (see also \cite{Gundlach:2017tqq}). In the specific case of $w=1/10$, significant fine-tuning of initial conditions may be required to observe deviations from the scaling law, as discussed in \cite{Celestino:2018ptx}. 
Nevertheless, it may only affect PBH masses well below the peak of the mass function, 
and we assume that the deviation from the scaling law is not relevant.}. 
Therefore, in this work, we follow the results of \cite{Baumgarte:2015aza,Celestino:2018ptx} and we reasonably assume that the PBH mass in the critical regime\footnote{Notice that the scaling PBH mass deviates for very large amplitudes far away from the critical value, as explicitly shown and quantified in \cite{Escriva:2019nsa} for a Gaussian profile} including non-spherical effects of $e\neq0$
is given by 
\begin{equation}
\label{eq:mass_spectrum}
M = M_H(\mu) \mathcal{K}(\mu-\mu_{\rm c}(e))^{\gamma}, 
\end{equation}
where $\mu_{\rm c}(e)$ is the function satisfying $e=e_{\rm c}(\mu_c(e))$. 
In addition, since 
relatively small values of $e$ are relevant to the PBH abundance, 
we have assumed that the $e$ dependences of $\mathcal K$ and $\gamma$ can be ignored, and they are fixed as in spherical collapse once the value of $w$ is given (see Ref.~\cite{Escriva:2021pmf} for the profile dependence of $\mathcal K$ and Ref.~\cite{Maison:1995cc} for the value of $\gamma$ as a function of $w$). $M_H$ is the mass of the cosmological horizon at the time $t_H$ of horizon reentry, which can be expressed assuming spherical symmetry as $M_H(\mu) = M_{k_p}(r_m e^{\bar{\zeta}_{sp}(r_m;\mu)})^{3(1+w)/(1+3w)}$, derived from the condition $a(t_H)H(t_H)r_m e^{-\bar{\zeta}_{sp}(r_m;\mu)} = 1$
with $r_m$ being the relevant length-scale of the fluctuation defined as $\bar{\zeta}'_{\rm sp}(r_m)+r_m \bar{\zeta}''_{\rm sp}(r_m) =0$. The mass of the horizon at the reentry of the wave mode $k_p$ in solar masses is given by $M_{k_p} = (g_{\star}/106.75)^{-1/6} \cdot 10^{20} \left(k_p/(1.56 \cdot 10^{13} \textrm{Mpc}^{-1})\right)^{-3(1+w)/(1+3w)} M_{\odot}$.

In Fig.~\ref{fig:non_spherical_mass}, one can see that the PBH mass vanishes on the line given by $\mu=\mu_{\rm c}(e)$. Therefore, under the assumption \eqref{eq:mass_spectrum}, 
the PBH mass depends on not only $\mu$ but also the ellipticity $e$, 
and a large value of $\mu$ can contribute to the abundance of the small mass PBH with $e$ very close to $e_{\rm c}(\mu)$. This behaviour is essential for the total PBH mass function which will be given later. 
 


\newsec{PBH number density}
According to (A.12) in \cite{1986ApJ...304...15B},  the peak number density distribution for the parameters $(\nu,\xi,e,p)$ 
is given by 
\begin{align}
\label{eq:picos}
    &\mathcal{N}_{\rm pk}(\nu, \xi , e, p ) \dd{ \nu} \dd{\xi} \dd{e} \dd{p}= \\ \nonumber
    & = \frac{5^{5/2} 3^{1/2}}{(2 \pi)^{3}} \left( \frac{\sigma_2}{\sigma_1}\right)^{3} \frac{\exp{-\bar{Q}}}{\sqrt{1-\gamma^2}} \xi^6 W(e,p)
    \dd{ \nu} \dd{\xi} \dd{e} \dd{p}, \\ \label{eq:F}
    & W(e,p) = (1-2p)\left[ (1+p)^2 - (3 e )^2 \right] e (e^2-p^2)\chi(e,p) ,  \\
    & \bar{Q} = \frac{\nu^2}{2} + \frac{(\xi-\xi_{*})^2}{2(1-\gamma^2)}+\frac{5}{2}(3e^2+p^2)\xi^2, 
\end{align}
where $\nu=\mu / \sigma_0$ and $\xi = \left.\nabla^2 \zeta\right|_{r=0} / \sigma_2$ are the variables that characterize the peak profile together with $e$ and $p$. 
The statistical parameters $\sigma_n$ are defined as $\sigma^2_n= \int k^{2n}\mathcal{P}_{\zeta}(k)d \ln k$ and $\xi_{*}=\gamma \nu$ with $\gamma= \sigma^2_1 /(\sigma_0 \, \sigma_2)$. 
The function $\chi(e,p)$ restricts the relevant domain as  
\begin{equation}
\label{eq:domain}
\left\{
\begin{aligned}
\chi &= 1 , \, \, \,  \, 0 \leq e \leq 1/4 , -e \leq p \leq e, \\
\chi &= 1 , \, \, \,  \, 1/4 \leq e \leq 1/2 , -(1-3e) \leq p \leq e, \\
\chi &= 0 , \, \, \,  \, \textrm{otherwise}. 
\end{aligned}
\right.
\end{equation}
If we integrate $\mathcal{N}_{\rm pk}(\nu, \xi, e, p )$ over $e$ and $p$, and divide Eq.\eqref{eq:picos} by it,  
we obtain the conditional probability $\mathcal{P}_{\rm e,p}(e,p \mid \nu, \xi )$ for the parameters $e$ and $p$ with the constraint that a peak is characterized by $\nu$ and $\xi$ as (see (C.4) in \cite{1986ApJ...304...15B})
%
\begin{equation}
    \mathcal{P}_{\rm e,p}dedp = \frac{3^{2} 5^{5/2}}{\sqrt{2\pi}} \frac{\xi^8}{f(\xi)}\exp{-\frac{5}{2} \xi^2 (3 e^2 + p^2)}W(e,p)dedp, 
\label{eq:prob_e_p}
\end{equation}
where $f(\xi)$ is
\begin{align}
    &f(\xi)=\frac{1}{2}\xi(\xi^2-3)\left(\erf\left[\frac{1}{2}\sqrt{\frac{5}{2}}\xi\right]+\erf \left[ \sqrt{\frac{5}{2}}\xi\right]\right) \nonumber\\ 
    &+\sqrt{\frac{2}{5\pi}}\left[\mathcal{C}_1(\xi)\exp{-\frac{5}{8}\xi^2}+\mathcal{C}_2(\xi)\exp{-\frac{5}{2}\xi^2}\right]
\label{eq:f}
\end{align}
with 
\begin{align}
     \mathcal{C}_1(\xi) = \frac{8}{5}+\frac{31}{4}\xi^2  \ ,  \, \, \, \,  \mathcal{C}_{2}(\xi) = -\frac{8}{5}+\frac{1}{2}\xi^2.  \nonumber
\end{align}
Equation \eqref{eq:prob_e_p} quantifies the likelihood of different configurations of the typical profile as functions of $e$ and $p$. In Fig.~\ref{fig:P_ep}, we present the distribution for $\xi = \nu = 8$. The figure shows that for large peaks, $\nu \gg 1$, the most likely configuration is sharply concentrated at mean value with small $e$ and $p$ (see also \cite{companion}).
Since the probability distribution is peaked around $p\approx0$, we may neglect the $p$ dependence of the threshold.

Following \cite{Yoo:2018kvb,Yoo:2020dkz}, the height of the peak $\nu$ can be rewritten in terms of the PBH mass through \eqref{eq:mass_spectrum} with the Jacobian $|J|(M,e)=\partial \mu/ \partial \ln M$. 
Then, the number density of the PBH mass $M$ is given by integrating over all the other variables as
%
\begin{equation}
    n_{\rm PBH}(M)  =  \left( \int \mathcal{N}_{\rm pk}(\nu(M,e), \xi , e, p )   \abs{J}^{-1} d\xi de dp \right)  /\sigma_0.
\end{equation}
%
%
%
%
Considering a monochromatic spectrum, we have $\sigma_{n} = \sigma_0 k^{n}$, which implies $\gamma = 1$. 
We can then integrate Eq.~\eqref{eq:picos} in the limit $\gamma \rightarrow 1$, using the identity $\textrm{lim}_{\epsilon \rightarrow 0} \, \exp{-(x-x_{\star})^2/(2\epsilon^2)} / \epsilon = \sqrt{2\pi} \delta(x-x_{\star})$, which involves a Dirac delta function in the integration, leading to $\xi = \xi_{\star} = \nu$. Thus we obtain
\begin{align}
\nonumber
&n_{\rm PBH}(M)= \int de dp  \sqrt{\frac{3}{\mathcal{A}_{\zeta}}}\left(\frac{5}{2 \pi} \right)^{5/2} \nu(M,e)^8 W(e,p)  \\ 
& \times \exp{-\frac{\nu(M,e)^2}{2}(1+15e^2+5p^2)}  \abs{J}^{-1}(M,e)\,,   
\label{eq:final_abundance}
\\ \nonumber 
 &\abs{J}(M,e) \equiv   \bigg|\frac{\partial \ln M}{\partial \mu}\bigg| 
 \\ 
 &=   \bigg| \frac{3(1+w)}{1+3w} \left(\frac{d \bar{\zeta}_{sp}(r_m;\mu)}{d \mu}\right)_{\mu=\mu(M,e)} 
+\frac{\gamma}{\mu(M,e)-\mu_{\rm c}(e)} \bigg|. 
\label{eq:jacobian}
\end{align}
We can make an extra step by performing the integration for $p$ 
to write Eq.\eqref{eq:final_abundance} as
\begin{align}
n_{\rm PBH}(M) &= \int_{e=0}^{e=1/4} g(M, e) h_{1}(\nu(M,e),e) de
\nonumber \\ 
&+ \int_{e=1/4}^{e=1/2}g(M, e) h_{2}(\nu(M,e),e) de, 
\label{eq:simplified_fpbh}
\end{align}
where $g(M, e)$ is given by
\begin{align}
\label{eq:final_abundance2}
 &g(M, e)= \sqrt{\frac{3}{\mathcal{A}_{\zeta}}}\left(\frac{5}{2 \pi} \right)^{5/2} 
 \nonumber \\
&\times \exp{-\frac{\nu(M, e)^2}{2}(1+15e^2)}  \nu(M, e)^8 \abs{J}(M,e), 
\end{align}
and the functions $h_{1}$ and $h_{2}$ are shown in the appendix \ref{sec:appendix}, which have been computed taking into account the integration domain $\chi(e,p)$ of Eq.~\eqref{eq:domain}. 
In the case of the absence of non-spherical effects, i.e., $e=p=0$, we recover the result of Ref.~\cite{Yoo:2018kvb}.

Finally, we can define the mass function of PBHs as the ratio of the energy density of PBHs to the total dark matter with a mass $M$ in the mass range $[M, M e ^{d \ln M}]$ (logarithmic bin), which reads as
\begin{equation}
    f_{\rm PBH}(M) d \ln M = \frac{M n_{\rm PBH}(M)}{\rho_{\rm DM}} d\ln M, 
\end{equation}
where $\rho_{\rm DM} = 3 M^{2}_{\rm pl} H^{2}_{0} \, \Omega_{\rm DM}$.  

\newsec{Numerical results}

\begin{figure} 
    \centering
    \includegraphics[width=0.45\textwidth]{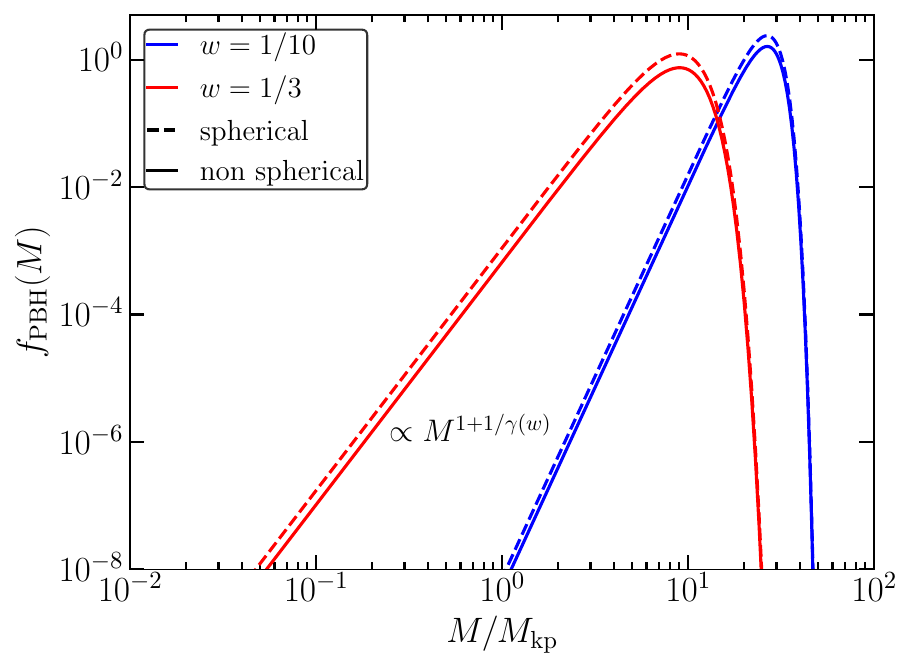}
    \caption{PBH mass function for the case $k_p$ in terms of $M_{k_p}$. The dashed line denotes the 
    result
    assuming spherical symmetry, 
    and 
    the solid line shows the estimation taking into account the non-spherical configurations.} 
    \label{fig:non_sphericalpbh}
\end{figure}

We consider a power spectrum $\mathcal{P}_{\zeta}$ that is peaked at the scale $k_p = 10^{13.5}, \textrm{Mpc}^{-1}$ and set $g_{*}=106.75$, which results in the mass function being peaked within the 
asteroid mass range.\footnote{Choosing different values of $k_p$ does not alter our conclusions regarding the non-spherical effects on the PBH mass function, so we focus on a single $k_p$ scale.} 
We first compute the mass function in the spherical case (with $e=p=0$) to determine the value of $\mathcal{A}_{\zeta}$ that yields a total PBH fraction compatible with all dark matter, $f_{\rm PBH}^{\rm tot} = \int f_{\rm PBH}(M) d\ln M = 1$. 
Using the numerical results from \cite{Escriva:2019nsa}, we obtain $\mathcal{A} \approx 1.242 \cdot 10^{-3}$, $\nu_c \approx 8.8$, and $M_{k_p} \approx 8.36 \times 10^{-15}M_{\odot}$ for $w=1/10$, and $\mathcal{A} \approx 4.870 \cdot 10^{-3}$, $\nu_c \approx 8.7$, and $M_{k_p} \approx 1.22 \times 10^{-14}M_{\odot}$ for $w=1/3$. 
We note that, using the typical profile $\bar{\zeta}_{sp}$ of Eq.~\eqref{eq:non_spherical_zeta}, 
from \cite{Escriva:2019nsa}, we obtain $\mu_{\rm c,sp} \approx 0.3095$ and $\mu_{\rm c,sp}\approx 0.6061$ for $w=1/10$ and $w=1/3$, respectively with $\mathcal{K} \sim 4$ \cite{Escriva:2021pmf}. 
%
We use these values to compute the PBH mass function, including non-spherical effects. 
First, we fix the values of $M/M_{k_p}$, and 
numerically solve Eq.~\eqref{eq:mass_spectrum} for $\mu$
as a function of $M$ and $e$ introducing the dependence $\mu_{\rm c}(e)$.  
For the given function $\mu(M,e)$, 
we make a spline interpolation. 
We then introduce this spline in the integration~\eqref{eq:simplified_fpbh}, 
and repeat the process for each mass ratio $M/M_{k_p}$. The results are shown in Fig.\ref{fig:non_sphericalpbh}. Comparing the spherical case (dashed line) and the case accounting for non-spherical effects (solid line), 
we find the difference in the mass function is not significant with a factor of approximately $1-2$ for both of $w=1/3$ and $1/10$ cases. 
Our findings suggest that, 
for the tested scenario of a monochromatic power spectrum, 
non-spherical effects do not play a significant role 
both for a radiation fluid $w=1/3$ and for a moderately softer equation of state with $w=1/10$.

The low mass-tail is dominated by the behaviour $\sim M^{1+1/\gamma(w)}$, due to the Jacobian term (Eq.~\eqref{eq:jacobian}) in Eq.~\eqref{eq:final_abundance}. In that regime, non-spherical effects play a crucial role, i.e., a small deviation from sphericity will avoid fluctuation collapse into a black hole as discussed in \cite{companion}. 
Nevertheless, because of the assumption of the critical collapse in Eq.~\eqref{eq:mass_spectrum}, the mass function will be dominated by fluctuations that are very close to its critical $\mu_{\rm c}(e)$,  
near the boundary of the line given by 
$e= \mathcal{K}_e [(\mu-\mu_{\rm c,sp})/\mu_{\rm c,sp}]^{ \gamma_e}$ as we shown in Fig.~\ref{fig:non_spherical_mass} for small $M/M_{k_p}$.


\newsec{Conclusions and discusion}

In this letter, we have used the results from $3+1$ relativistic numerical simulations \cite{companion} and peak theory \cite{1986ApJ...304...15B} to estimate the impact of non-spherical, ellipsoidal configurations on the mass function of PBHs, focusing on a monochromatic power spectrum. 
Specifically, we apply peak theory for obtaining $f_{\rm PBH}(M)$ with a novel approach that incorporates non-spherical effects
with a plausible assumption of the ellipticity dependent mass scaling \eqref{eq:mass_spectrum}. 


Our results show that the effects of non-sphericity on the mass function of PBHs are 
insignificant
both for the radiation-dominated case ($w=1/3$) and for a softer equation of state ($w=1/10$). 
This indicates that the reduction in threshold values found in \cite{companion} is not significant enough to influence the overall abundance of PBHs, demonstrating that non-spherical effects are not impactful. This differs from the conclusion in \cite{Kuhnel:2016exn}, as our approach incorporates numerical results alongside Eq.\eqref{eq:prob_e_p}, which accounts for the conditional probability of $e$ and $p$ in terms of the peak height. This contrasts with the Press-Schechter approach, where all non-spherical configurations are assumed to be equally likely. The primary implication of our findings is that, for the tested case of a monochromatic power spectrum, the assumption of spherical symmetry in calculating the PBH abundance and mass function is highly accurate, leading to modifications in the mass function by a factor of less than $1-2$. This conclusion also extends to softer equations of state, where non-spherical effects could have been expected to play a significant role. 

Moreover, although non-spherical effects are important when fluctuations are near the critical value (hindering them from collapsing into black holes \cite{companion}), due to the critical scaling law for PBH mass, the low-mass tail behaviour of the mass function remains largely unaffected. 
This is because, 
for each ellipticity $e$, the PBH 
mass function
is dominated again by fluctuations 
with the amplitude $\mu$ very close to $\mu_{\rm c}(e)$, 
and the same exponent $f_{\rm PBH}(M) \sim M^{1+1/\gamma}$ is preserved
associated with the collapse of the perfect fluid.

In this work, we have employed a specific curvature profile derived from a monochromatic power spectrum. 
It would be interesting to test our results with alternative profiles. 
Specifically, it would be worthwhile to explore whether the functional form of $\mu_{\rm c}(e)$ could induce significant differences in the mass function when considering other spectra (in particular broad ones) or distribution of profiles. 
Additionally, while our analysis has focused on Gaussian fluctuations, 
investigating non-Gaussian fluctuations would be insightful, as both the curvature profile and the probability distribution of $e$ and $p$ would change. Moreover, obtaining an exact PBH mass function requires determining the PBH mass for various configurations of $(e,p)$ in the critical regime and verifying whether the scaling regime holds for arbitrary values of $\mu - \mu_{\rm c}(e,p)$. 
In our work, we assumed that the mass 
scaling 
remains unaffected by nonspherical configurations, as we may 
expect from \cite{Gundlach:1997vy,Baumgarte:2015aza,Celestino:2018ptx}. 
However, substantial changes in the mass scaling law
with respect to ellipticity ($e$) and prolateness ($p$) could affect the shape of the PBH mass function, $f_{\rm PBH}(M)$, especially its slope in the low-mass regime. 
A numerical investigation would be valuable to 
find how large asphericity is needed in our specific setting for PBH formation 
to break the mass scaling suggested in \cite{Gundlach:1997vy,Baumgarte:2015aza,Celestino:2018ptx} 
(see for instance \cite{Choptuik:2003ac,Marouda:2024epb}, where large deviations from sphericity can modify the scaling regime in the scenario of collapse of axisymmetric scalar fields).

We leave these questions for future research. Further investigation would be desirable to determine whether non-spherical effects could have a more substantial impact than what we have demonstrated here. 


\noindent\textbf{Acknowledgments.} 
A.E. acknowledges support from the JSPS Postdoctoral Fellowships for Research in Japan (Graduate School of Sciences, Nagoya University). C.Y. is supported in part by JSPS KAKENHI Grant Nos. 20H05850 and 20H05853.

\onecolumngrid
\appendix

\section{Appendix: Analytical formulas}
\label{sec:appendix}

Here, we explicitly provide the analytical functions $h_1,h_2$ used for the determination of the number density of peaks in Eq.(14).

\begin{equation*}
h_1(\nu,e)= -\frac{e}{125 \nu^5} \left[\sqrt{10 \pi } \left(25 e^2 \left(9 e^2-1\right) \nu ^4+\left(5-30 e^2\right) \nu ^2-9\right) \erf\left(\sqrt{\frac{5}{2}} e \nu \right)+10 e \, \nu \,  \exp{-\frac{5}{2} e^2
   \nu ^2} \left(5 \left(9 e^2-1\right) \nu ^2+9\right)\right], 
\end{equation*}

\begin{align*}
\nonumber
h_2(\nu, e)&=-\frac{e}{250 \nu ^6}\Biggl[\sqrt{10 \pi } \nu  \left(25 e^2 \left(9 e^2-1\right) \nu ^4+\left(5-30 e^2\right) \nu ^2-9\right) \erf \left(\sqrt{\frac{5}{2}} (1-3 e) \nu \right)\\ \nonumber
&+\sqrt{10 \pi } \nu 
   \left(25 e^2 \left(9 e^2-1\right) \nu ^4+\left(5-30 e^2\right) \nu ^2-9\right) \erf \left(\sqrt{\frac{5}{2}} e \nu \right)\\ \nonumber
&-2 \exp{-\frac{5}{2} \left(9 e^2+1\right) \nu ^2}
   \Biggl(\exp{15 e \nu ^2} \left(25 e^2 (3 e+1) \nu ^4+5 \left(32 e^2-21 e-1\right) \nu ^2+16\right)+\\
&+\exp{\frac{5}{2} \left(8 e^2+1\right) \nu ^2} \left(-225 e^3 \nu ^4+160 e^2 \nu ^2+5
   e \left(5 \nu ^2-9\right) \nu ^2-16\right)\Biggr)\Biggr].  
\end{align*}


\bibliographystyle{apsrev4-1}
\bibliography{bibfile}

\begin{thebibliography}{61}%
\makeatletter
\providecommand \@ifxundefined [1]{%
 \@ifx{#1\undefined}
}%
\providecommand \@ifnum [1]{%
 \ifnum #1\expandafter \@firstoftwo
 \else \expandafter \@secondoftwo
 \fi
}%
\providecommand \@ifx [1]{%
 \ifx #1\expandafter \@firstoftwo
 \else \expandafter \@secondoftwo
 \fi
}%
\providecommand \natexlab [1]{#1}%
\providecommand \enquote  [1]{``#1''}%
\providecommand \bibnamefont  [1]{#1}%
\providecommand \bibfnamefont [1]{#1}%
\providecommand \citenamefont [1]{#1}%
\providecommand \href@noop [0]{\@secondoftwo}%
\providecommand \href [0]{\begingroup \@sanitize@url \@href}%
\providecommand \@href[1]{\@@startlink{#1}\@@href}%
\providecommand \@@href[1]{\endgroup#1\@@endlink}%
\providecommand \@sanitize@url [0]{\catcode `\\12\catcode `\$12\catcode `\&12\catcode `\#12\catcode `\^12\catcode `\_12\catcode `\%12\relax}%
\providecommand \@@startlink[1]{}%
\providecommand \@@endlink[0]{}%
\providecommand \url  [0]{\begingroup\@sanitize@url \@url }%
\providecommand \@url [1]{\endgroup\@href {#1}{\urlprefix }}%
\providecommand \urlprefix  [0]{URL }%
\providecommand \Eprint [0]{\href }%
\providecommand \doibase [0]{http://dx.doi.org/}%
\providecommand \selectlanguage [0]{\@gobble}%
\providecommand \bibinfo  [0]{\@secondoftwo}%
\providecommand \bibfield  [0]{\@secondoftwo}%
\providecommand \translation [1]{[#1]}%
\providecommand \BibitemOpen [0]{}%
\providecommand \bibitemStop [0]{}%
\providecommand \bibitemNoStop [0]{.\EOS\space}%
\providecommand \EOS [0]{\spacefactor3000\relax}%
\providecommand \BibitemShut  [1]{\csname bibitem#1\endcsname}%
\let\auto@bib@innerbib\@empty
\bibitem [{\citenamefont {Escriv\`a}\ and\ \citenamefont {Yoo}(2025)}]{companion}%
  \BibitemOpen
  \bibfield  {author} {\bibinfo {author} {\bibfnamefont {A.}~\bibnamefont {Escriv\`a}}\ and\ \bibinfo {author} {\bibfnamefont {C.-M.}\ \bibnamefont {Yoo}},\ }\href {\doibase 10.1103/PhysRevD.112.083518} {\bibfield  {journal} {\bibinfo  {journal} {Phys. Rev. D}\ }\textbf {\bibinfo {volume} {112}},\ \bibinfo {pages} {083518} (\bibinfo {year} {2025})}\BibitemShut {NoStop}%
\bibitem [{\citenamefont {Zel'dovich}\ and\ \citenamefont {Novikov}(1967)}]{Zeldovich:1967lct}%
  \BibitemOpen
  \bibfield  {author} {\bibinfo {author} {\bibfnamefont {Y.~B.}\ \bibnamefont {Zel'dovich}}\ and\ \bibinfo {author} {\bibfnamefont {I.~D.}\ \bibnamefont {Novikov}},\ }\href@noop {} {\bibfield  {journal} {\bibinfo  {journal} {Sov. Astron.}\ }\textbf {\bibinfo {volume} {10}},\ \bibinfo {pages} {602} (\bibinfo {year} {1967})}\BibitemShut {NoStop}%
\bibitem [{\citenamefont {Hawking}(1971)}]{10.1093/mnras/152.1.75}%
  \BibitemOpen
  \bibfield  {author} {\bibinfo {author} {\bibfnamefont {S.}~\bibnamefont {Hawking}},\ }\href {\doibase 10.1093/mnras/152.1.75} {\bibfield  {journal} {\bibinfo  {journal} {Monthly Notices of the Royal Astronomical Society}\ }\textbf {\bibinfo {volume} {152}},\ \bibinfo {pages} {75} (\bibinfo {year} {1971})},\ \Eprint {http://arxiv.org/abs/https://academic.oup.com/mnras/article-pdf/152/1/75/9360899/mnras152-0075.pdf} {https://academic.oup.com/mnras/article-pdf/152/1/75/9360899/mnras152-0075.pdf} \BibitemShut {NoStop}%
\bibitem [{\citenamefont {Carr}\ and\ \citenamefont {Hawking}(1974)}]{Carr:1974nx}%
  \BibitemOpen
  \bibfield  {author} {\bibinfo {author} {\bibfnamefont {B.~J.}\ \bibnamefont {Carr}}\ and\ \bibinfo {author} {\bibfnamefont {S.~W.}\ \bibnamefont {Hawking}},\ }\href {\doibase 10.1093/mnras/168.2.399} {\bibfield  {journal} {\bibinfo  {journal} {Mon. Not. Roy. Astron. Soc.}\ }\textbf {\bibinfo {volume} {168}},\ \bibinfo {pages} {399} (\bibinfo {year} {1974})}\BibitemShut {NoStop}%
\bibitem [{\citenamefont {Escriv\`a}\ \emph {et~al.}(2022)\citenamefont {Escriv\`a}, \citenamefont {Kuhnel},\ and\ \citenamefont {Tada}}]{Escriva:2022duf}%
  \BibitemOpen
  \bibfield  {author} {\bibinfo {author} {\bibfnamefont {A.}~\bibnamefont {Escriv\`a}}, \bibinfo {author} {\bibfnamefont {F.}~\bibnamefont {Kuhnel}}, \ and\ \bibinfo {author} {\bibfnamefont {Y.}~\bibnamefont {Tada}},\ }\href {\doibase 10.1016/B978-0-32-395636-9.00012-8} {\  (\bibinfo {year} {2022}),\ 10.1016/B978-0-32-395636-9.00012-8},\ \Eprint {http://arxiv.org/abs/2211.05767} {arXiv:2211.05767 [astro-ph.CO]} \BibitemShut {NoStop}%
\bibitem [{\citenamefont {Khlopov}(2010)}]{Khlopov:2008qy}%
  \BibitemOpen
  \bibfield  {author} {\bibinfo {author} {\bibfnamefont {M.~Y.}\ \bibnamefont {Khlopov}},\ }\href {\doibase 10.1088/1674-4527/10/6/001} {\bibfield  {journal} {\bibinfo  {journal} {Res. Astron. Astrophys.}\ }\textbf {\bibinfo {volume} {10}},\ \bibinfo {pages} {495} (\bibinfo {year} {2010})},\ \Eprint {http://arxiv.org/abs/0801.0116} {arXiv:0801.0116 [astro-ph]} \BibitemShut {NoStop}%
\bibitem [{\citenamefont {Sasaki}\ \emph {et~al.}(2018)\citenamefont {Sasaki}, \citenamefont {Suyama}, \citenamefont {Tanaka},\ and\ \citenamefont {Yokoyama}}]{Sasaki:2018dmp}%
  \BibitemOpen
  \bibfield  {author} {\bibinfo {author} {\bibfnamefont {M.}~\bibnamefont {Sasaki}}, \bibinfo {author} {\bibfnamefont {T.}~\bibnamefont {Suyama}}, \bibinfo {author} {\bibfnamefont {T.}~\bibnamefont {Tanaka}}, \ and\ \bibinfo {author} {\bibfnamefont {S.}~\bibnamefont {Yokoyama}},\ }\href {\doibase 10.1088/1361-6382/aaa7b4} {\bibfield  {journal} {\bibinfo  {journal} {Class. Quant. Grav.}\ }\textbf {\bibinfo {volume} {35}},\ \bibinfo {pages} {063001} (\bibinfo {year} {2018})},\ \Eprint {http://arxiv.org/abs/1801.05235} {arXiv:1801.05235 [astro-ph.CO]} \BibitemShut {NoStop}%
\bibitem [{\citenamefont {Carr}\ \emph {et~al.}(2021)\citenamefont {Carr}, \citenamefont {Kohri}, \citenamefont {Sendouda},\ and\ \citenamefont {Yokoyama}}]{Carr:2020gox}%
  \BibitemOpen
  \bibfield  {author} {\bibinfo {author} {\bibfnamefont {B.}~\bibnamefont {Carr}}, \bibinfo {author} {\bibfnamefont {K.}~\bibnamefont {Kohri}}, \bibinfo {author} {\bibfnamefont {Y.}~\bibnamefont {Sendouda}}, \ and\ \bibinfo {author} {\bibfnamefont {J.}~\bibnamefont {Yokoyama}},\ }\href {\doibase 10.1088/1361-6633/ac1e31} {\bibfield  {journal} {\bibinfo  {journal} {Rept. Prog. Phys.}\ }\textbf {\bibinfo {volume} {84}},\ \bibinfo {pages} {116902} (\bibinfo {year} {2021})},\ \Eprint {http://arxiv.org/abs/2002.12778} {arXiv:2002.12778 [astro-ph.CO]} \BibitemShut {NoStop}%
\bibitem [{\citenamefont {Green}\ and\ \citenamefont {Kavanagh}(2021)}]{Green:2020jor}%
  \BibitemOpen
  \bibfield  {author} {\bibinfo {author} {\bibfnamefont {A.~M.}\ \bibnamefont {Green}}\ and\ \bibinfo {author} {\bibfnamefont {B.~J.}\ \bibnamefont {Kavanagh}},\ }\href {\doibase 10.1088/1361-6471/abc534} {\bibfield  {journal} {\bibinfo  {journal} {J. Phys. G}\ }\textbf {\bibinfo {volume} {48}},\ \bibinfo {pages} {043001} (\bibinfo {year} {2021})},\ \Eprint {http://arxiv.org/abs/2007.10722} {arXiv:2007.10722 [astro-ph.CO]} \BibitemShut {NoStop}%
\bibitem [{\citenamefont {Carr}\ and\ \citenamefont {Kuhnel}(2020)}]{Carr:2020xqk}%
  \BibitemOpen
  \bibfield  {author} {\bibinfo {author} {\bibfnamefont {B.}~\bibnamefont {Carr}}\ and\ \bibinfo {author} {\bibfnamefont {F.}~\bibnamefont {Kuhnel}},\ }\href {\doibase 10.1146/annurev-nucl-050520-125911} {\bibfield  {journal} {\bibinfo  {journal} {Ann. Rev. Nucl. Part. Sci.}\ }\textbf {\bibinfo {volume} {70}},\ \bibinfo {pages} {355} (\bibinfo {year} {2020})},\ \Eprint {http://arxiv.org/abs/2006.02838} {arXiv:2006.02838 [astro-ph.CO]} \BibitemShut {NoStop}%
\bibitem [{\citenamefont {Carr}(1975)}]{Carr:1975qj}%
  \BibitemOpen
  \bibfield  {author} {\bibinfo {author} {\bibfnamefont {B.~J.}\ \bibnamefont {Carr}},\ }\href {\doibase 10.1086/153853} {\bibfield  {journal} {\bibinfo  {journal} {Astrophys. J.}\ }\textbf {\bibinfo {volume} {201}},\ \bibinfo {pages} {1} (\bibinfo {year} {1975})}\BibitemShut {NoStop}%
\bibitem [{\citenamefont {Chapline}(1975)}]{Chapline:1975ojl}%
  \BibitemOpen
  \bibfield  {author} {\bibinfo {author} {\bibfnamefont {G.~F.}\ \bibnamefont {Chapline}},\ }\href {\doibase 10.1038/253251a0} {\bibfield  {journal} {\bibinfo  {journal} {Nature}\ }\textbf {\bibinfo {volume} {253}},\ \bibinfo {pages} {251} (\bibinfo {year} {1975})}\BibitemShut {NoStop}%
\bibitem [{\citenamefont {Escriv\`a}(2022)}]{Escriva:2021aeh}%
  \BibitemOpen
  \bibfield  {author} {\bibinfo {author} {\bibfnamefont {A.}~\bibnamefont {Escriv\`a}},\ }\href {\doibase 10.3390/universe8020066} {\bibfield  {journal} {\bibinfo  {journal} {Universe}\ }\textbf {\bibinfo {volume} {8}},\ \bibinfo {pages} {66} (\bibinfo {year} {2022})},\ \Eprint {http://arxiv.org/abs/2111.12693} {arXiv:2111.12693 [gr-qc]} \BibitemShut {NoStop}%
\bibitem [{\citenamefont {{Bardeen}}\ \emph {et~al.}(1986)\citenamefont {{Bardeen}}, \citenamefont {{Bond}}, \citenamefont {{Kaiser}},\ and\ \citenamefont {{Szalay}}}]{1986ApJ...304...15B}%
  \BibitemOpen
  \bibfield  {author} {\bibinfo {author} {\bibfnamefont {J.~M.}\ \bibnamefont {{Bardeen}}}, \bibinfo {author} {\bibfnamefont {J.~R.}\ \bibnamefont {{Bond}}}, \bibinfo {author} {\bibfnamefont {N.}~\bibnamefont {{Kaiser}}}, \ and\ \bibinfo {author} {\bibfnamefont {A.~S.}\ \bibnamefont {{Szalay}}},\ }\href {\doibase 10.1086/164143} {\bibfield  {journal} {\bibinfo  {journal} {\apj}\ }\textbf {\bibinfo {volume} {304}},\ \bibinfo {pages} {15} (\bibinfo {year} {1986})}\BibitemShut {NoStop}%
\bibitem [{\citenamefont {Mizuguchi}\ \emph {et~al.}(2024)\citenamefont {Mizuguchi}, \citenamefont {Murata},\ and\ \citenamefont {Tada}}]{Mizuguchi:2024kbl}%
  \BibitemOpen
  \bibfield  {author} {\bibinfo {author} {\bibfnamefont {Y.}~\bibnamefont {Mizuguchi}}, \bibinfo {author} {\bibfnamefont {T.}~\bibnamefont {Murata}}, \ and\ \bibinfo {author} {\bibfnamefont {Y.}~\bibnamefont {Tada}},\ }\href {\doibase 10.1088/1475-7516/2024/12/050} {\bibfield  {journal} {\bibinfo  {journal} {JCAP}\ }\textbf {\bibinfo {volume} {12}},\ \bibinfo {pages} {050} (\bibinfo {year} {2024})},\ \Eprint {http://arxiv.org/abs/2405.10692} {arXiv:2405.10692 [astro-ph.CO]} \BibitemShut {NoStop}%
\bibitem [{\citenamefont {{Lin}}\ \emph {et~al.}(1965)\citenamefont {{Lin}}, \citenamefont {{Mestel}},\ and\ \citenamefont {{Shu}}}]{1965ApJ...142.1431L}%
  \BibitemOpen
  \bibfield  {author} {\bibinfo {author} {\bibfnamefont {C.~C.}\ \bibnamefont {{Lin}}}, \bibinfo {author} {\bibfnamefont {L.}~\bibnamefont {{Mestel}}}, \ and\ \bibinfo {author} {\bibfnamefont {F.~H.}\ \bibnamefont {{Shu}}},\ }\href {\doibase 10.1086/148428} {\bibfield  {journal} {\bibinfo  {journal} {\apj}\ }\textbf {\bibinfo {volume} {142}},\ \bibinfo {pages} {1431} (\bibinfo {year} {1965})}\BibitemShut {NoStop}%
\bibitem [{\citenamefont {Khlopov}\ and\ \citenamefont {Polnarev}(1980)}]{KHLOPOV1980383}%
  \BibitemOpen
  \bibfield  {author} {\bibinfo {author} {\bibfnamefont {M.}~\bibnamefont {Khlopov}}\ and\ \bibinfo {author} {\bibfnamefont {A.}~\bibnamefont {Polnarev}},\ }\href {\doibase https://doi.org/10.1016/0370-2693(80)90624-3} {\bibfield  {journal} {\bibinfo  {journal} {Physics Letters B}\ }\textbf {\bibinfo {volume} {97}},\ \bibinfo {pages} {383} (\bibinfo {year} {1980})}\BibitemShut {NoStop}%
\bibitem [{\citenamefont {Yoo}(2022)}]{Yoo:2022mzl}%
  \BibitemOpen
  \bibfield  {author} {\bibinfo {author} {\bibfnamefont {C.-M.}\ \bibnamefont {Yoo}},\ }\href {\doibase 10.3390/galaxies10060112} {\bibfield  {journal} {\bibinfo  {journal} {Galaxies}\ }\textbf {\bibinfo {volume} {10}},\ \bibinfo {pages} {112} (\bibinfo {year} {2022})},\ \Eprint {http://arxiv.org/abs/2211.13512} {arXiv:2211.13512 [astro-ph.CO]} \BibitemShut {NoStop}%
\bibitem [{\citenamefont {Germani}\ and\ \citenamefont {Sheth}(2023)}]{Germani:2023ojx}%
  \BibitemOpen
  \bibfield  {author} {\bibinfo {author} {\bibfnamefont {C.}~\bibnamefont {Germani}}\ and\ \bibinfo {author} {\bibfnamefont {R.~K.}\ \bibnamefont {Sheth}},\ }\href {\doibase 10.3390/universe9090421} {\bibfield  {journal} {\bibinfo  {journal} {Universe}\ }\textbf {\bibinfo {volume} {9}},\ \bibinfo {pages} {421} (\bibinfo {year} {2023})},\ \Eprint {http://arxiv.org/abs/2308.02971} {arXiv:2308.02971 [astro-ph.CO]} \BibitemShut {NoStop}%
\bibitem [{\citenamefont {Young}(2024)}]{Young:2024jsu}%
  \BibitemOpen
  \bibfield  {author} {\bibinfo {author} {\bibfnamefont {S.}~\bibnamefont {Young}},\ }\href@noop {} {\  (\bibinfo {year} {2024})},\ \Eprint {http://arxiv.org/abs/2405.13259} {arXiv:2405.13259 [astro-ph.CO]} \BibitemShut {NoStop}%
\bibitem [{\citenamefont {Shibata}\ and\ \citenamefont {Sasaki}(1999)}]{Shibata:1999zs}%
  \BibitemOpen
  \bibfield  {author} {\bibinfo {author} {\bibfnamefont {M.}~\bibnamefont {Shibata}}\ and\ \bibinfo {author} {\bibfnamefont {M.}~\bibnamefont {Sasaki}},\ }\href {\doibase 10.1103/PhysRevD.60.084002} {\bibfield  {journal} {\bibinfo  {journal} {Phys. Rev. D}\ }\textbf {\bibinfo {volume} {60}},\ \bibinfo {pages} {084002} (\bibinfo {year} {1999})},\ \Eprint {http://arxiv.org/abs/gr-qc/9905064} {arXiv:gr-qc/9905064} \BibitemShut {NoStop}%
\bibitem [{\citenamefont {Niemeyer}\ and\ \citenamefont {Jedamzik}(1999)}]{Niemeyer:1999ak}%
  \BibitemOpen
  \bibfield  {author} {\bibinfo {author} {\bibfnamefont {J.~C.}\ \bibnamefont {Niemeyer}}\ and\ \bibinfo {author} {\bibfnamefont {K.}~\bibnamefont {Jedamzik}},\ }\href {\doibase 10.1103/PhysRevD.59.124013} {\bibfield  {journal} {\bibinfo  {journal} {Phys. Rev. D}\ }\textbf {\bibinfo {volume} {59}},\ \bibinfo {pages} {124013} (\bibinfo {year} {1999})},\ \Eprint {http://arxiv.org/abs/astro-ph/9901292} {arXiv:astro-ph/9901292} \BibitemShut {NoStop}%
\bibitem [{\citenamefont {Musco}\ \emph {et~al.}(2005)\citenamefont {Musco}, \citenamefont {Miller},\ and\ \citenamefont {Rezzolla}}]{Musco:2004ak}%
  \BibitemOpen
  \bibfield  {author} {\bibinfo {author} {\bibfnamefont {I.}~\bibnamefont {Musco}}, \bibinfo {author} {\bibfnamefont {J.~C.}\ \bibnamefont {Miller}}, \ and\ \bibinfo {author} {\bibfnamefont {L.}~\bibnamefont {Rezzolla}},\ }\href {\doibase 10.1088/0264-9381/22/7/013} {\bibfield  {journal} {\bibinfo  {journal} {Class. Quant. Grav.}\ }\textbf {\bibinfo {volume} {22}},\ \bibinfo {pages} {1405} (\bibinfo {year} {2005})},\ \Eprint {http://arxiv.org/abs/gr-qc/0412063} {arXiv:gr-qc/0412063} \BibitemShut {NoStop}%
\bibitem [{\citenamefont {Musco}\ \emph {et~al.}(2009)\citenamefont {Musco}, \citenamefont {Miller},\ and\ \citenamefont {Polnarev}}]{Musco:2008hv}%
  \BibitemOpen
  \bibfield  {author} {\bibinfo {author} {\bibfnamefont {I.}~\bibnamefont {Musco}}, \bibinfo {author} {\bibfnamefont {J.~C.}\ \bibnamefont {Miller}}, \ and\ \bibinfo {author} {\bibfnamefont {A.~G.}\ \bibnamefont {Polnarev}},\ }\href {\doibase 10.1088/0264-9381/26/23/235001} {\bibfield  {journal} {\bibinfo  {journal} {Class. Quant. Grav.}\ }\textbf {\bibinfo {volume} {26}},\ \bibinfo {pages} {235001} (\bibinfo {year} {2009})},\ \Eprint {http://arxiv.org/abs/0811.1452} {arXiv:0811.1452 [gr-qc]} \BibitemShut {NoStop}%
\bibitem [{\citenamefont {Nakama}\ \emph {et~al.}(2014)\citenamefont {Nakama}, \citenamefont {Harada}, \citenamefont {Polnarev},\ and\ \citenamefont {Yokoyama}}]{Nakama:2013ica}%
  \BibitemOpen
  \bibfield  {author} {\bibinfo {author} {\bibfnamefont {T.}~\bibnamefont {Nakama}}, \bibinfo {author} {\bibfnamefont {T.}~\bibnamefont {Harada}}, \bibinfo {author} {\bibfnamefont {A.~G.}\ \bibnamefont {Polnarev}}, \ and\ \bibinfo {author} {\bibfnamefont {J.}~\bibnamefont {Yokoyama}},\ }\href {\doibase 10.1088/1475-7516/2014/01/037} {\bibfield  {journal} {\bibinfo  {journal} {JCAP}\ }\textbf {\bibinfo {volume} {01}},\ \bibinfo {pages} {037} (\bibinfo {year} {2014})},\ \Eprint {http://arxiv.org/abs/1310.3007} {arXiv:1310.3007 [gr-qc]} \BibitemShut {NoStop}%
\bibitem [{\citenamefont {Nakama}(2014)}]{Nakama:2014fra}%
  \BibitemOpen
  \bibfield  {author} {\bibinfo {author} {\bibfnamefont {T.}~\bibnamefont {Nakama}},\ }\href {\doibase 10.1088/1475-7516/2014/10/040} {\bibfield  {journal} {\bibinfo  {journal} {JCAP}\ }\textbf {\bibinfo {volume} {10}},\ \bibinfo {pages} {040} (\bibinfo {year} {2014})},\ \Eprint {http://arxiv.org/abs/1408.0955} {arXiv:1408.0955 [gr-qc]} \BibitemShut {NoStop}%
\bibitem [{\citenamefont {Harada}\ \emph {et~al.}(2015)\citenamefont {Harada}, \citenamefont {Yoo}, \citenamefont {Nakama},\ and\ \citenamefont {Koga}}]{Harada:2015yda}%
  \BibitemOpen
  \bibfield  {author} {\bibinfo {author} {\bibfnamefont {T.}~\bibnamefont {Harada}}, \bibinfo {author} {\bibfnamefont {C.-M.}\ \bibnamefont {Yoo}}, \bibinfo {author} {\bibfnamefont {T.}~\bibnamefont {Nakama}}, \ and\ \bibinfo {author} {\bibfnamefont {Y.}~\bibnamefont {Koga}},\ }\href {\doibase 10.1103/PhysRevD.91.084057} {\bibfield  {journal} {\bibinfo  {journal} {Phys. Rev. D}\ }\textbf {\bibinfo {volume} {91}},\ \bibinfo {pages} {084057} (\bibinfo {year} {2015})},\ \Eprint {http://arxiv.org/abs/1503.03934} {arXiv:1503.03934 [gr-qc]} \BibitemShut {NoStop}%
\bibitem [{\citenamefont {Escriv\`a}(2020)}]{Escriva:2019nsa}%
  \BibitemOpen
  \bibfield  {author} {\bibinfo {author} {\bibfnamefont {A.}~\bibnamefont {Escriv\`a}},\ }\href {\doibase 10.1016/j.dark.2020.100466} {\bibfield  {journal} {\bibinfo  {journal} {Phys. Dark Univ.}\ }\textbf {\bibinfo {volume} {27}},\ \bibinfo {pages} {100466} (\bibinfo {year} {2020})},\ \Eprint {http://arxiv.org/abs/1907.13065} {arXiv:1907.13065 [gr-qc]} \BibitemShut {NoStop}%
\bibitem [{\citenamefont {Musco}(2019)}]{Musco:2018rwt}%
  \BibitemOpen
  \bibfield  {author} {\bibinfo {author} {\bibfnamefont {I.}~\bibnamefont {Musco}},\ }\href {\doibase 10.1103/PhysRevD.100.123524} {\bibfield  {journal} {\bibinfo  {journal} {Phys. Rev. D}\ }\textbf {\bibinfo {volume} {100}},\ \bibinfo {pages} {123524} (\bibinfo {year} {2019})},\ \Eprint {http://arxiv.org/abs/1809.02127} {arXiv:1809.02127 [gr-qc]} \BibitemShut {NoStop}%
\bibitem [{\citenamefont {Escriv\`a}\ \emph {et~al.}(2020)\citenamefont {Escriv\`a}, \citenamefont {Germani},\ and\ \citenamefont {Sheth}}]{Escriva:2019phb}%
  \BibitemOpen
  \bibfield  {author} {\bibinfo {author} {\bibfnamefont {A.}~\bibnamefont {Escriv\`a}}, \bibinfo {author} {\bibfnamefont {C.}~\bibnamefont {Germani}}, \ and\ \bibinfo {author} {\bibfnamefont {R.~K.}\ \bibnamefont {Sheth}},\ }\href {\doibase 10.1103/PhysRevD.101.044022} {\bibfield  {journal} {\bibinfo  {journal} {Phys. Rev. D}\ }\textbf {\bibinfo {volume} {101}},\ \bibinfo {pages} {044022} (\bibinfo {year} {2020})},\ \Eprint {http://arxiv.org/abs/1907.13311} {arXiv:1907.13311 [gr-qc]} \BibitemShut {NoStop}%
\bibitem [{\citenamefont {Escriv\`a}\ \emph {et~al.}(2021)\citenamefont {Escriv\`a}, \citenamefont {Germani},\ and\ \citenamefont {Sheth}}]{Escriva:2020tak}%
  \BibitemOpen
  \bibfield  {author} {\bibinfo {author} {\bibfnamefont {A.}~\bibnamefont {Escriv\`a}}, \bibinfo {author} {\bibfnamefont {C.}~\bibnamefont {Germani}}, \ and\ \bibinfo {author} {\bibfnamefont {R.~K.}\ \bibnamefont {Sheth}},\ }\href {\doibase 10.1088/1475-7516/2021/01/030} {\bibfield  {journal} {\bibinfo  {journal} {JCAP}\ }\textbf {\bibinfo {volume} {01}},\ \bibinfo {pages} {030} (\bibinfo {year} {2021})},\ \Eprint {http://arxiv.org/abs/2007.05564} {arXiv:2007.05564 [gr-qc]} \BibitemShut {NoStop}%
\bibitem [{\citenamefont {Yoo}\ \emph {et~al.}(2022)\citenamefont {Yoo}, \citenamefont {Harada}, \citenamefont {Hirano}, \citenamefont {Okawa},\ and\ \citenamefont {Sasaki}}]{Yoo:2021fxs}%
  \BibitemOpen
  \bibfield  {author} {\bibinfo {author} {\bibfnamefont {C.-M.}\ \bibnamefont {Yoo}}, \bibinfo {author} {\bibfnamefont {T.}~\bibnamefont {Harada}}, \bibinfo {author} {\bibfnamefont {S.}~\bibnamefont {Hirano}}, \bibinfo {author} {\bibfnamefont {H.}~\bibnamefont {Okawa}}, \ and\ \bibinfo {author} {\bibfnamefont {M.}~\bibnamefont {Sasaki}},\ }\href {\doibase 10.1103/PhysRevD.105.103538} {\bibfield  {journal} {\bibinfo  {journal} {Phys. Rev. D}\ }\textbf {\bibinfo {volume} {105}},\ \bibinfo {pages} {103538} (\bibinfo {year} {2022})},\ \Eprint {http://arxiv.org/abs/2112.12335} {arXiv:2112.12335 [gr-qc]} \BibitemShut {NoStop}%
\bibitem [{\citenamefont {Escriv\`a}\ and\ \citenamefont {Subils}(2023)}]{Escriva:2022yaf}%
  \BibitemOpen
  \bibfield  {author} {\bibinfo {author} {\bibfnamefont {A.}~\bibnamefont {Escriv\`a}}\ and\ \bibinfo {author} {\bibfnamefont {J.~G.}\ \bibnamefont {Subils}},\ }\href {\doibase 10.1103/PhysRevD.107.L041301} {\bibfield  {journal} {\bibinfo  {journal} {Phys. Rev. D}\ }\textbf {\bibinfo {volume} {107}},\ \bibinfo {pages} {L041301} (\bibinfo {year} {2023})},\ \Eprint {http://arxiv.org/abs/2211.15674} {arXiv:2211.15674 [astro-ph.CO]} \BibitemShut {NoStop}%
\bibitem [{\citenamefont {Escriv\`a}\ and\ \citenamefont {Yoo}(2024)}]{Escriva:2023qnq}%
  \BibitemOpen
  \bibfield  {author} {\bibinfo {author} {\bibfnamefont {A.}~\bibnamefont {Escriv\`a}}\ and\ \bibinfo {author} {\bibfnamefont {C.-M.}\ \bibnamefont {Yoo}},\ }\href {\doibase 10.1088/1475-7516/2024/04/048} {\bibfield  {journal} {\bibinfo  {journal} {JCAP}\ }\textbf {\bibinfo {volume} {04}},\ \bibinfo {pages} {048} (\bibinfo {year} {2024})},\ \Eprint {http://arxiv.org/abs/2310.16482} {arXiv:2310.16482 [gr-qc]} \BibitemShut {NoStop}%
\bibitem [{\citenamefont {Uehara}\ \emph {et~al.}(2024)\citenamefont {Uehara}, \citenamefont {Escriv\`a}, \citenamefont {Harada}, \citenamefont {Saito},\ and\ \citenamefont {Yoo}}]{Uehara:2024yyp}%
  \BibitemOpen
  \bibfield  {author} {\bibinfo {author} {\bibfnamefont {K.}~\bibnamefont {Uehara}}, \bibinfo {author} {\bibfnamefont {A.}~\bibnamefont {Escriv\`a}}, \bibinfo {author} {\bibfnamefont {T.}~\bibnamefont {Harada}}, \bibinfo {author} {\bibfnamefont {D.}~\bibnamefont {Saito}}, \ and\ \bibinfo {author} {\bibfnamefont {C.-M.}\ \bibnamefont {Yoo}},\ }\href@noop {} {\  (\bibinfo {year} {2024})},\ \Eprint {http://arxiv.org/abs/2401.06329} {arXiv:2401.06329 [gr-qc]} \BibitemShut {NoStop}%
\bibitem [{\citenamefont {Yoo}\ \emph {et~al.}(2020)\citenamefont {Yoo}, \citenamefont {Harada},\ and\ \citenamefont {Okawa}}]{Yoo:2020lmg}%
  \BibitemOpen
  \bibfield  {author} {\bibinfo {author} {\bibfnamefont {C.-M.}\ \bibnamefont {Yoo}}, \bibinfo {author} {\bibfnamefont {T.}~\bibnamefont {Harada}}, \ and\ \bibinfo {author} {\bibfnamefont {H.}~\bibnamefont {Okawa}},\ }\href {\doibase 10.1103/PhysRevD.102.043526} {\bibfield  {journal} {\bibinfo  {journal} {Phys. Rev. D}\ }\textbf {\bibinfo {volume} {102}},\ \bibinfo {pages} {043526} (\bibinfo {year} {2020})},\ \bibinfo {note} {[Erratum: Phys.Rev.D 107, 049901 (2023)]},\ \Eprint {http://arxiv.org/abs/2004.01042} {arXiv:2004.01042 [gr-qc]} \BibitemShut {NoStop}%
\bibitem [{\citenamefont {Choptuik}(1993)}]{Choptuik:1992jv}%
  \BibitemOpen
  \bibfield  {author} {\bibinfo {author} {\bibfnamefont {M.~W.}\ \bibnamefont {Choptuik}},\ }\href {\doibase 10.1103/PhysRevLett.70.9} {\bibfield  {journal} {\bibinfo  {journal} {Phys. Rev. Lett.}\ }\textbf {\bibinfo {volume} {70}},\ \bibinfo {pages} {9} (\bibinfo {year} {1993})}\BibitemShut {NoStop}%
\bibitem [{\citenamefont {Koike}\ \emph {et~al.}(1995)\citenamefont {Koike}, \citenamefont {Hara},\ and\ \citenamefont {Adachi}}]{Koike:1995jm}%
  \BibitemOpen
  \bibfield  {author} {\bibinfo {author} {\bibfnamefont {T.}~\bibnamefont {Koike}}, \bibinfo {author} {\bibfnamefont {T.}~\bibnamefont {Hara}}, \ and\ \bibinfo {author} {\bibfnamefont {S.}~\bibnamefont {Adachi}},\ }\href {\doibase 10.1103/PhysRevLett.74.5170} {\bibfield  {journal} {\bibinfo  {journal} {Phys. Rev. Lett.}\ }\textbf {\bibinfo {volume} {74}},\ \bibinfo {pages} {5170} (\bibinfo {year} {1995})},\ \Eprint {http://arxiv.org/abs/gr-qc/9503007} {arXiv:gr-qc/9503007} \BibitemShut {NoStop}%
\bibitem [{\citenamefont {Maison}(1996)}]{Maison:1995cc}%
  \BibitemOpen
  \bibfield  {author} {\bibinfo {author} {\bibfnamefont {D.}~\bibnamefont {Maison}},\ }\href {\doibase 10.1016/0370-2693(95)01381-4} {\bibfield  {journal} {\bibinfo  {journal} {Phys. Lett. B}\ }\textbf {\bibinfo {volume} {366}},\ \bibinfo {pages} {82} (\bibinfo {year} {1996})},\ \Eprint {http://arxiv.org/abs/gr-qc/9504008} {arXiv:gr-qc/9504008} \BibitemShut {NoStop}%
\bibitem [{\citenamefont {Hawke}\ and\ \citenamefont {Stewart}(2002)}]{IHawke_2002}%
  \BibitemOpen
  \bibfield  {author} {\bibinfo {author} {\bibfnamefont {I.}~\bibnamefont {Hawke}}\ and\ \bibinfo {author} {\bibfnamefont {J.~M.}\ \bibnamefont {Stewart}},\ }\href {\doibase 10.1088/0264-9381/19/14/310} {\bibfield  {journal} {\bibinfo  {journal} {Classical and Quantum Gravity}\ }\textbf {\bibinfo {volume} {19}},\ \bibinfo {pages} {3687} (\bibinfo {year} {2002})}\BibitemShut {NoStop}%
\bibitem [{\citenamefont {Niemeyer}\ and\ \citenamefont {Jedamzik}(1998)}]{Niemeyer:1997mt}%
  \BibitemOpen
  \bibfield  {author} {\bibinfo {author} {\bibfnamefont {J.~C.}\ \bibnamefont {Niemeyer}}\ and\ \bibinfo {author} {\bibfnamefont {K.}~\bibnamefont {Jedamzik}},\ }\href {\doibase 10.1103/PhysRevLett.80.5481} {\bibfield  {journal} {\bibinfo  {journal} {Phys. Rev. Lett.}\ }\textbf {\bibinfo {volume} {80}},\ \bibinfo {pages} {5481} (\bibinfo {year} {1998})},\ \Eprint {http://arxiv.org/abs/astro-ph/9709072} {arXiv:astro-ph/9709072} \BibitemShut {NoStop}%
\bibitem [{\citenamefont {Yokoyama}(1998)}]{Yokoyama:1998xd}%
  \BibitemOpen
  \bibfield  {author} {\bibinfo {author} {\bibfnamefont {J.}~\bibnamefont {Yokoyama}},\ }\href {\doibase 10.1103/PhysRevD.58.107502} {\bibfield  {journal} {\bibinfo  {journal} {Phys. Rev. D}\ }\textbf {\bibinfo {volume} {58}},\ \bibinfo {pages} {107502} (\bibinfo {year} {1998})},\ \Eprint {http://arxiv.org/abs/gr-qc/9804041} {arXiv:gr-qc/9804041} \BibitemShut {NoStop}%
\bibitem [{\citenamefont {Kitajima}\ \emph {et~al.}(2021)\citenamefont {Kitajima}, \citenamefont {Tada}, \citenamefont {Yokoyama},\ and\ \citenamefont {Yoo}}]{Kitajima:2021fpq}%
  \BibitemOpen
  \bibfield  {author} {\bibinfo {author} {\bibfnamefont {N.}~\bibnamefont {Kitajima}}, \bibinfo {author} {\bibfnamefont {Y.}~\bibnamefont {Tada}}, \bibinfo {author} {\bibfnamefont {S.}~\bibnamefont {Yokoyama}}, \ and\ \bibinfo {author} {\bibfnamefont {C.-M.}\ \bibnamefont {Yoo}},\ }\href {\doibase 10.1088/1475-7516/2021/10/053} {\bibfield  {journal} {\bibinfo  {journal} {JCAP}\ }\textbf {\bibinfo {volume} {10}},\ \bibinfo {pages} {053} (\bibinfo {year} {2021})},\ \Eprint {http://arxiv.org/abs/2109.00791} {arXiv:2109.00791 [astro-ph.CO]} \BibitemShut {NoStop}%
\bibitem [{\citenamefont {Gundlach}(2002)}]{Gundlach:1999cw}%
  \BibitemOpen
  \bibfield  {author} {\bibinfo {author} {\bibfnamefont {C.}~\bibnamefont {Gundlach}},\ }\href {\doibase 10.1103/PhysRevD.65.084021} {\bibfield  {journal} {\bibinfo  {journal} {Phys. Rev. D}\ }\textbf {\bibinfo {volume} {65}},\ \bibinfo {pages} {084021} (\bibinfo {year} {2002})},\ \Eprint {http://arxiv.org/abs/gr-qc/9906124} {arXiv:gr-qc/9906124} \BibitemShut {NoStop}%
\bibitem [{\citenamefont {Baumgarte}\ and\ \citenamefont {Montero}(2015)}]{Baumgarte:2015aza}%
  \BibitemOpen
  \bibfield  {author} {\bibinfo {author} {\bibfnamefont {T.~W.}\ \bibnamefont {Baumgarte}}\ and\ \bibinfo {author} {\bibfnamefont {P.~J.}\ \bibnamefont {Montero}},\ }\href {\doibase 10.1103/PhysRevD.92.124065} {\bibfield  {journal} {\bibinfo  {journal} {Phys. Rev. D}\ }\textbf {\bibinfo {volume} {92}},\ \bibinfo {pages} {124065} (\bibinfo {year} {2015})},\ \Eprint {http://arxiv.org/abs/1509.08730} {arXiv:1509.08730 [gr-qc]} \BibitemShut {NoStop}%
\bibitem [{\citenamefont {Celestino}\ and\ \citenamefont {Baumgarte}(2018)}]{Celestino:2018ptx}%
  \BibitemOpen
  \bibfield  {author} {\bibinfo {author} {\bibfnamefont {J.}~\bibnamefont {Celestino}}\ and\ \bibinfo {author} {\bibfnamefont {T.~W.}\ \bibnamefont {Baumgarte}},\ }\href {\doibase 10.1103/PhysRevD.98.024053} {\bibfield  {journal} {\bibinfo  {journal} {Phys. Rev. D}\ }\textbf {\bibinfo {volume} {98}},\ \bibinfo {pages} {024053} (\bibinfo {year} {2018})},\ \Eprint {http://arxiv.org/abs/1805.10442} {arXiv:1805.10442 [gr-qc]} \BibitemShut {NoStop}%
\bibitem [{\citenamefont {Chiba}\ and\ \citenamefont {Yokoyama}(2017)}]{Chiba:2017rvs}%
  \BibitemOpen
  \bibfield  {author} {\bibinfo {author} {\bibfnamefont {T.}~\bibnamefont {Chiba}}\ and\ \bibinfo {author} {\bibfnamefont {S.}~\bibnamefont {Yokoyama}},\ }\href {\doibase 10.1093/ptep/ptx087} {\bibfield  {journal} {\bibinfo  {journal} {PTEP}\ }\textbf {\bibinfo {volume} {2017}},\ \bibinfo {pages} {083E01} (\bibinfo {year} {2017})},\ \Eprint {http://arxiv.org/abs/1704.06573} {arXiv:1704.06573 [gr-qc]} \BibitemShut {NoStop}%
\bibitem [{\citenamefont {K\"uhnel}\ and\ \citenamefont {Sandstad}(2016)}]{Kuhnel:2016exn}%
  \BibitemOpen
  \bibfield  {author} {\bibinfo {author} {\bibfnamefont {F.}~\bibnamefont {K\"uhnel}}\ and\ \bibinfo {author} {\bibfnamefont {M.}~\bibnamefont {Sandstad}},\ }\href {\doibase 10.1103/PhysRevD.94.063514} {\bibfield  {journal} {\bibinfo  {journal} {Phys. Rev. D}\ }\textbf {\bibinfo {volume} {94}},\ \bibinfo {pages} {063514} (\bibinfo {year} {2016})},\ \Eprint {http://arxiv.org/abs/1602.04815} {arXiv:1602.04815 [astro-ph.CO]} \BibitemShut {NoStop}%
\bibitem [{\citenamefont {Harada}\ \emph {et~al.}(2016)\citenamefont {Harada}, \citenamefont {Yoo}, \citenamefont {Kohri}, \citenamefont {Nakao},\ and\ \citenamefont {Jhingan}}]{Harada:2016mhb}%
  \BibitemOpen
  \bibfield  {author} {\bibinfo {author} {\bibfnamefont {T.}~\bibnamefont {Harada}}, \bibinfo {author} {\bibfnamefont {C.-M.}\ \bibnamefont {Yoo}}, \bibinfo {author} {\bibfnamefont {K.}~\bibnamefont {Kohri}}, \bibinfo {author} {\bibfnamefont {K.-i.}\ \bibnamefont {Nakao}}, \ and\ \bibinfo {author} {\bibfnamefont {S.}~\bibnamefont {Jhingan}},\ }\href {\doibase 10.3847/1538-4357/833/1/61} {\bibfield  {journal} {\bibinfo  {journal} {Astrophys. J.}\ }\textbf {\bibinfo {volume} {833}},\ \bibinfo {pages} {61} (\bibinfo {year} {2016})},\ \Eprint {http://arxiv.org/abs/1609.01588} {arXiv:1609.01588 [astro-ph.CO]} \BibitemShut {NoStop}%
\bibitem [{\citenamefont {{Press}}\ and\ \citenamefont {{Schechter}}(1974)}]{1974ApJ...187..425P}%
  \BibitemOpen
  \bibfield  {author} {\bibinfo {author} {\bibfnamefont {W.~H.}\ \bibnamefont {{Press}}}\ and\ \bibinfo {author} {\bibfnamefont {P.}~\bibnamefont {{Schechter}}},\ }\href {\doibase 10.1086/152650} {\bibfield  {journal} {\bibinfo  {journal} {\apj}\ }\textbf {\bibinfo {volume} {187}},\ \bibinfo {pages} {425} (\bibinfo {year} {1974})}\BibitemShut {NoStop}%
\bibitem [{\citenamefont {Sheth}\ \emph {et~al.}(2001)\citenamefont {Sheth}, \citenamefont {Mo},\ and\ \citenamefont {Tormen}}]{Sheth:1999su}%
  \BibitemOpen
  \bibfield  {author} {\bibinfo {author} {\bibfnamefont {R.~K.}\ \bibnamefont {Sheth}}, \bibinfo {author} {\bibfnamefont {H.~J.}\ \bibnamefont {Mo}}, \ and\ \bibinfo {author} {\bibfnamefont {G.}~\bibnamefont {Tormen}},\ }\href {\doibase 10.1046/j.1365-8711.2001.04006.x} {\bibfield  {journal} {\bibinfo  {journal} {Mon. Not. Roy. Astron. Soc.}\ }\textbf {\bibinfo {volume} {323}},\ \bibinfo {pages} {1} (\bibinfo {year} {2001})},\ \Eprint {http://arxiv.org/abs/astro-ph/9907024} {arXiv:astro-ph/9907024} \BibitemShut {NoStop}%
\bibitem [{\citenamefont {Angrick}\ and\ \citenamefont {Bartelmann}(2010)}]{Angrick:2010qg}%
  \BibitemOpen
  \bibfield  {author} {\bibinfo {author} {\bibfnamefont {C.}~\bibnamefont {Angrick}}\ and\ \bibinfo {author} {\bibfnamefont {M.}~\bibnamefont {Bartelmann}},\ }\href {\doibase 10.1051/0004-6361/201014147} {\bibfield  {journal} {\bibinfo  {journal} {Astron. Astrophys.}\ }\textbf {\bibinfo {volume} {518}},\ \bibinfo {pages} {A38} (\bibinfo {year} {2010})},\ \Eprint {http://arxiv.org/abs/1001.4984} {arXiv:1001.4984 [astro-ph.CO]} \BibitemShut {NoStop}%
\bibitem [{\citenamefont {Baumgarte}\ and\ \citenamefont {Gundlach}(2016)}]{Baumgarte:2016xjw}%
  \BibitemOpen
  \bibfield  {author} {\bibinfo {author} {\bibfnamefont {T.~W.}\ \bibnamefont {Baumgarte}}\ and\ \bibinfo {author} {\bibfnamefont {C.}~\bibnamefont {Gundlach}},\ }\href {\doibase 10.1103/PhysRevLett.116.221103} {\bibfield  {journal} {\bibinfo  {journal} {Phys. Rev. Lett.}\ }\textbf {\bibinfo {volume} {116}},\ \bibinfo {pages} {221103} (\bibinfo {year} {2016})},\ \Eprint {http://arxiv.org/abs/1603.04373} {arXiv:1603.04373 [gr-qc]} \BibitemShut {NoStop}%
\bibitem [{\citenamefont {Gundlach}\ and\ \citenamefont {Baumgarte}(2016)}]{Gundlach:2016jzm}%
  \BibitemOpen
  \bibfield  {author} {\bibinfo {author} {\bibfnamefont {C.}~\bibnamefont {Gundlach}}\ and\ \bibinfo {author} {\bibfnamefont {T.~W.}\ \bibnamefont {Baumgarte}},\ }\href {\doibase 10.1103/PhysRevD.94.084012} {\bibfield  {journal} {\bibinfo  {journal} {Phys. Rev. D}\ }\textbf {\bibinfo {volume} {94}},\ \bibinfo {pages} {084012} (\bibinfo {year} {2016})},\ \Eprint {http://arxiv.org/abs/1608.00491} {arXiv:1608.00491 [gr-qc]} \BibitemShut {NoStop}%
\bibitem [{\citenamefont {Gundlach}\ and\ \citenamefont {Baumgarte}(2018)}]{Gundlach:2017tqq}%
  \BibitemOpen
  \bibfield  {author} {\bibinfo {author} {\bibfnamefont {C.}~\bibnamefont {Gundlach}}\ and\ \bibinfo {author} {\bibfnamefont {T.~W.}\ \bibnamefont {Baumgarte}},\ }\href {\doibase 10.1103/PhysRevD.97.064006} {\bibfield  {journal} {\bibinfo  {journal} {Phys. Rev. D}\ }\textbf {\bibinfo {volume} {97}},\ \bibinfo {pages} {064006} (\bibinfo {year} {2018})},\ \Eprint {http://arxiv.org/abs/1712.05741} {arXiv:1712.05741 [gr-qc]} \BibitemShut {NoStop}%
\bibitem [{\citenamefont {Escriv\`a}\ and\ \citenamefont {Romano}(2021)}]{Escriva:2021pmf}%
  \BibitemOpen
  \bibfield  {author} {\bibinfo {author} {\bibfnamefont {A.}~\bibnamefont {Escriv\`a}}\ and\ \bibinfo {author} {\bibfnamefont {A.~E.}\ \bibnamefont {Romano}},\ }\href {\doibase 10.1088/1475-7516/2021/05/066} {\bibfield  {journal} {\bibinfo  {journal} {JCAP}\ }\textbf {\bibinfo {volume} {05}},\ \bibinfo {pages} {066} (\bibinfo {year} {2021})},\ \Eprint {http://arxiv.org/abs/2103.03867} {arXiv:2103.03867 [gr-qc]} \BibitemShut {NoStop}%
\bibitem [{\citenamefont {Yoo}\ \emph {et~al.}(2018)\citenamefont {Yoo}, \citenamefont {Harada}, \citenamefont {Garriga},\ and\ \citenamefont {Kohri}}]{Yoo:2018kvb}%
  \BibitemOpen
  \bibfield  {author} {\bibinfo {author} {\bibfnamefont {C.-M.}\ \bibnamefont {Yoo}}, \bibinfo {author} {\bibfnamefont {T.}~\bibnamefont {Harada}}, \bibinfo {author} {\bibfnamefont {J.}~\bibnamefont {Garriga}}, \ and\ \bibinfo {author} {\bibfnamefont {K.}~\bibnamefont {Kohri}},\ }\href {\doibase 10.1093/ptep/pty120} {\bibfield  {journal} {\bibinfo  {journal} {PTEP}\ }\textbf {\bibinfo {volume} {2018}},\ \bibinfo {pages} {123E01} (\bibinfo {year} {2018})},\ \bibinfo {note} {[Erratum: PTEP 2024, 049202 (2024)]},\ \Eprint {http://arxiv.org/abs/1805.03946} {arXiv:1805.03946 [astro-ph.CO]} \BibitemShut {NoStop}%
\bibitem [{\citenamefont {Yoo}\ \emph {et~al.}(2021)\citenamefont {Yoo}, \citenamefont {Harada}, \citenamefont {Hirano},\ and\ \citenamefont {Kohri}}]{Yoo:2020dkz}%
  \BibitemOpen
  \bibfield  {author} {\bibinfo {author} {\bibfnamefont {C.-M.}\ \bibnamefont {Yoo}}, \bibinfo {author} {\bibfnamefont {T.}~\bibnamefont {Harada}}, \bibinfo {author} {\bibfnamefont {S.}~\bibnamefont {Hirano}}, \ and\ \bibinfo {author} {\bibfnamefont {K.}~\bibnamefont {Kohri}},\ }\href {\doibase 10.1093/ptep/ptaa155} {\bibfield  {journal} {\bibinfo  {journal} {PTEP}\ }\textbf {\bibinfo {volume} {2021}},\ \bibinfo {pages} {013E02} (\bibinfo {year} {2021})},\ \bibinfo {note} {[Erratum: PTEP 2024, 049203 (2024)]},\ \Eprint {http://arxiv.org/abs/2008.02425} {arXiv:2008.02425 [astro-ph.CO]} \BibitemShut {NoStop}%
\bibitem [{\citenamefont {Gundlach}(1998)}]{Gundlach:1997vy}%
  \BibitemOpen
  \bibfield  {author} {\bibinfo {author} {\bibfnamefont {C.}~\bibnamefont {Gundlach}},\ }\href {\doibase 10.1103/PhysRevD.57.R7075} {\bibfield  {journal} {\bibinfo  {journal} {Phys. Rev. D}\ }\textbf {\bibinfo {volume} {57}},\ \bibinfo {pages} {7075} (\bibinfo {year} {1998})},\ \Eprint {http://arxiv.org/abs/gr-qc/9710066} {arXiv:gr-qc/9710066} \BibitemShut {NoStop}%
\bibitem [{\citenamefont {Choptuik}\ \emph {et~al.}(2003)\citenamefont {Choptuik}, \citenamefont {Hirschmann}, \citenamefont {Liebling},\ and\ \citenamefont {Pretorius}}]{Choptuik:2003ac}%
  \BibitemOpen
  \bibfield  {author} {\bibinfo {author} {\bibfnamefont {M.~W.}\ \bibnamefont {Choptuik}}, \bibinfo {author} {\bibfnamefont {E.~W.}\ \bibnamefont {Hirschmann}}, \bibinfo {author} {\bibfnamefont {S.~L.}\ \bibnamefont {Liebling}}, \ and\ \bibinfo {author} {\bibfnamefont {F.}~\bibnamefont {Pretorius}},\ }\href {\doibase 10.1103/PhysRevD.68.044007} {\bibfield  {journal} {\bibinfo  {journal} {Phys. Rev. D}\ }\textbf {\bibinfo {volume} {68}},\ \bibinfo {pages} {044007} (\bibinfo {year} {2003})},\ \Eprint {http://arxiv.org/abs/gr-qc/0305003} {arXiv:gr-qc/0305003} \BibitemShut {NoStop}%
\bibitem [{\citenamefont {Marouda}\ \emph {et~al.}(2024)\citenamefont {Marouda}, \citenamefont {Cors}, \citenamefont {R\"uter}, \citenamefont {Atteneder},\ and\ \citenamefont {Hilditch}}]{Marouda:2024epb}%
  \BibitemOpen
  \bibfield  {author} {\bibinfo {author} {\bibfnamefont {K.}~\bibnamefont {Marouda}}, \bibinfo {author} {\bibfnamefont {D.}~\bibnamefont {Cors}}, \bibinfo {author} {\bibfnamefont {H.~R.}\ \bibnamefont {R\"uter}}, \bibinfo {author} {\bibfnamefont {F.}~\bibnamefont {Atteneder}}, \ and\ \bibinfo {author} {\bibfnamefont {D.}~\bibnamefont {Hilditch}},\ }\href {\doibase 10.1103/PhysRevD.109.124042} {\bibfield  {journal} {\bibinfo  {journal} {Phys. Rev. D}\ }\textbf {\bibinfo {volume} {109}},\ \bibinfo {pages} {124042} (\bibinfo {year} {2024})},\ \Eprint {http://arxiv.org/abs/2402.06724} {arXiv:2402.06724 [gr-qc]} \BibitemShut {NoStop}%
\end{thebibliography}%
\end{document}